\documentclass{article}
\usepackage{a4wide}
\usepackage{cite}
\usepackage{amsmath}
\usepackage{amssymb}
\usepackage{epsfig}
\usepackage{subfigure}
\newcommand{\half}{{\frac{1}{2}}}
\newcommand{\qb}{{\overline{q}}}
\newcommand{\Hc}{{h_\mathrm{c}}}
\newcommand{\Hm}{{h_\mathrm{m}}}
\newcommand{\Jc}{{J_\mathrm{c}^{(r)}}}
\newcommand{\Jm}{{J_\mathrm{m}^{(r)}}}
\newcommand{\Ts}{{T_\mathrm{s}}}
\newcommand{\Td}{{T_\mathrm{d}}}
\newcommand{\Tat}{{T_\mathrm{AT}}}
\newcommand{\Tm}{{T_\mathrm{m}}}
\newcommand{\Tsz}{{T_\mathrm{s}^0}}
\newcommand{\Mm}{{M_\mathrm{min}}}
\newcommand{\he}{{h_\mathrm{eff}}}
\newcommand{\dFRS}{{F''_\mathrm{RS}(M)}}
\newcommand{\dFRSB}{{F''_\mathrm{RSB}(M)}}
\newcommand{\FFM}{{F_\mathrm{FM}}}
\newcommand{\FPM}{{F_\mathrm{PM}}}
\newcommand{\FSG}{{F_\mathrm{SG}}}
\newcommand{\FGFM}{{F_\mathrm{GFM}}}
\title{$p>2$ spin glasses with first order ferromagnetic transitions}
\author{Peter Gillin and David Sherrington\\
\small{Physics Department, University of Oxford,
Theoretical Physics, 1 Keble Road, Oxford. OX1 3NP}}
\date{}
\begin{document}
\maketitle
\begin{abstract}
\noindent
We consider an infinite-range spherical $p$-spin glass model with an
additional $r$-spin ferromagnetic interaction, both statically using a
replica analysis and dynamically via a generating functional
method. For $r>2$ we find that there are first order transitions to
ferromagnetic phases. For $r<p$ there are two ferromagnetic phases,
one non-glassy replica symmetric and one exhibiting glassy one-step
replica symmetry breaking and aging, whereas for $r \geqslant p$ only the
replica symmetric phase exists.
\end{abstract}
\section{Introduction}

The infinite-ranged spherical spin glass with $p>2$ body random
exchange interactions has attracted significant attention recently for
a number of reasons. Among these are (i)~it is exactly soluble in the
thermodynamic limit both for its equilibrium properties and for its
off-equilibrium macrodynamics (at least in the sense of coupled
equations for macroscopic order functions and \textsl{ans\"atze}
solving them in a long time limit), (ii)~it exhibits replica symmetry
breaking (RSB) and aging glassy macrostates, (iii)~the RSB is of the
one-step kind (1RSB) even down to the lowest temperatures. (In its
discontinuous variant, 1RSB is believed to be symptomatic of many
systems exhibiting glassy behaviour without Hamiltonian disorder.)

The original work \cite{cs:stat,chs:dyn} and most subsequent advances
have concentrated on situations where the exchange distribution is
symmetric, albeit possibly with an external field. More recently
\cite{hsn:r2} an extension was introduced to allow for an additional
ferromagnetic interaction, stimulated by the existence of many
physical systems with large coherently coordinated attractors, such as
most real spin glass materials with appropriate concentrations
\cite{mydosh:expt}, neural networks \cite{hkp:nnets}, proteins
\cite{wot:protfold}, and error-correcting codes\cite{s:ecc}. This
extension was however limited to two-body ferromagnetic exchange, with
corresponding second-order ferromagnetic transitions. In this paper we
extend the study further to include $r>2$ body ferromagnetic
interactions for which the onset of ferromagnetism is first
order. This is of relevance both for application to real systems (for
example, for error-correcting codes $r=p$) and since it brings in new
features (spinodal and thermodynamic transitions, metastability, and
suppression of glassy ferromagnetism).

\section{The model}

The Hamiltonian we use is
\begin{gather}
{\cal H} = \sum_{i_1 < i_2 \ldots < i_p} J_{i_1 \ldots i_p} \phi_{i_1}
\ldots \phi_{i_p} - \frac{J_0 (r-1)!}{N^{r-1}} \sum_{i_1 < i_2 \ldots <
i_r} \phi_{i_1} \ldots \phi_{i_r}
\label{ham}
\end{gather}
where the $J_{i_1 \ldots i_p}$ are independent quenched random
couplings given by a Gaussian distribution with zero mean and variance
\mbox{${p! J^2 / 2N^{p-1}}$}. The spins $\phi_i$ are real numbers
subject to the spherical constraint
\begin{gather}
\frac{1}{N} \sum_i \phi_i^2 = 1\:.
\label{spher}
\end{gather}
We consider \mbox{$p>2$} so that replica symmetry is broken for low
temperatures, spherical spins so that 1RSB is sufficient at all
temperatures, and infinite-ranged interactions so that mean field
theory is exact. For $r=1$ this reduces to the model of
\cite{cs:stat,chs:dyn}, for $r=2$ it becomes that of \cite{hsn:r2},
but we shall be interested also in $r>2$.

In considering the phase diagram, we will identify lines of four types. 
Since there are first order transitions, in the statics we must consider
the spinodal lines (where a phase appears) and the thermodynamic lines
(where it becomes thermodynamically preferred due to its free energy). A
modified replica analysis employing the criterion of marginal stability
(MS) leads to a different set of spinodal lines. Finally, there are
spinodal lines generated by a study of the Langevin dynamics at long
times. As in the cases studied previously, with an external field
\cite{chs:dyn} or a \mbox{2-spin} coupling \cite{hsn:r2}, the dynamic
lines are identical to the MS lines.

\section{Replica theory}

The equilibrium properties of this model are given by the replica
method \cite{mpv:book,s:rev} in which one-step replica symmetry
breaking is sufficient. This generates four order parameters governed
by self-consistent equations. Three of these describe the probability
distribution $P(q)$ of the overlap between pure states: $P(q)$ has two
$\delta$-function spikes, one corresponding to the self-overlap
\begin{align}
q_1 = q^{SS} &= \frac{1}{N} \left[ \sum_i \left( \langle \phi_i
\rangle^S \right)^2 \right]_\mathrm{av} \intertext{and one
corresponding to the mutual overlap} q_0 = q^{SS'} &= \frac{1}{N}
\left[ \sum_i \langle \phi_i \rangle^S \langle \phi_i \rangle^{S'}
\right]_\mathrm{av}, \qquad\qquad S \neq S'\,,
\end{align}
where $\langle \ldots \rangle ^S$ refers to the thermodynamic average
over the microstates of the pure state $S$ and $[\ldots]_\mathrm{av}$
to the average over the quenched disorder; the strength of these two
are \mbox{$1-x$} and $x$ respectively. The fourth order parameter is
the magnetization $M$.

The replica free energy $F$ is, in the thermodynamic limit ($N \to
\infty$), given by
\begin{gather}
\frac{2\beta F}{N} = \lim_{n \to 0} \frac{1}{n} \left[ -\frac{\beta^2
J^2}{2} \sum_{ab} q_{ab}^p - \frac{2\beta J_0}{r} \sum_a M_a^r - \ln\det
\left( q - M \otimes M \right) \right]\,,
\label{repFE}
\end{gather}
where the replica indices $a$ and $b$ run from $1$ to $n$, and $M
\otimes M$ represents the outer product of the vectors. Under the 1RSB
\textsl{ansatz} \cite{p:ansatz3} this becomes
\begin{gather}
\frac{2\beta F}{N} = - \frac{\beta^2 J^2}{2} (1 - \overline{q^p}) -
\frac{2 \beta J_0}{r} M^r - \ln(1-q_1) - \frac{1}{x} \ln \left(
\frac{1-\qb}{1-q_1} \right) - \frac{q_0 - M^2}{1-\qb}
\label{rsbFE}
\end{gather}
where we have defined an average over $P(q)$: for any value of $m$,
\begin{gather}
\overline{q^m} = (1-x) q_1^m + x q_0^m\:.
\end{gather}
Stationarity of $F$ leads to the self-consistent equations. Extremising
with respect to $M$ gives
\begin{align}
\beta J_0 M^{r-1} &= \frac{M}{1-\qb}\:;
\label{eqnM}
\intertext{with respect to $q_0$ and $q_1$ gives}
\half p \beta^2 J^2 q_0^{p-1} &= \frac{q_0 - M^2}{(1-\qb)^2}\,,
\label{eqn1q}
\\
\half p \beta^2 J^2 (q_1^{p-1} - q_0^{p-1}) &= \frac{q_1 - q_0} 
{(1-q_1)(1-\qb)}\:;
\label{eqn2q}
\end{align}
and with respect to $x$ gives
\begin{gather}
-\half \beta^2 J^2 (q_1^p - q_0^p) + \frac{1}{x^2} \ln \left( 
\frac{1-\qb}{1-q_1} \right) - \frac{q_1 - q_0}{x(1-\qb)} + \frac{(q_0 -
M^2)(q_1-q_0)}{(1-\qb)^2} = 0\:.
\label{eqnx}
\end{gather}
For the statics, equations (\ref{eqnM}), (\ref{eqn1q}), (\ref{eqn2q})
and (\ref{eqnx}) are solved for $M$, $q_0$, $q_1$, and $x$. We also
find it useful to consider the free energy with the magnetization
constrained; in this case (\ref{eqnM}) does not apply. For the
calculation under marginal stability, we abandon (\ref{eqnx}), and
instead insist that the lowest eigenvalue of the Hessian matrix of
(\ref{repFE}) in the $q_{ab}$ should vanish, which is to say that the
system should be marginally stable against small fluctuations in the
overlaps. The resulting equation is
\begin{gather}
\frac{1}{(1-q_1)^2} - \half p(p-1) \beta^2 J^2 q_1^{p-2} = 0\:.
\label{MS}
\end{gather}

\section{Dynamics}
The dynamics used are given by the Langevin equation
\begin{align}
\frac{\partial \phi_i}{\partial t} &= - \beta \frac{\partial {\cal
H}}{\partial \phi_i} - z(t) \phi_i(t) + \xi_i(t)\,,
\intertext{where $\xi_i(t)$ is a Gaussian thermal noise with zero mean and
satisfying}
\langle \xi_i(t) \xi_j(t') \rangle &= 2 \delta_{ij} \delta(t-t')\,,
\end{align}
and $z(t)$ is introduced to enforce the spherical constraint (\ref{spher}).
The standard generating function procedure \cite{sz:dyn,kt:sspsgdyn}
yields a self-consistent mean-field equation of motion:
\begin{align}
\frac{\partial \phi}{\partial t} &= \half p(p-1) \beta^2 J^2
\int_{-\infty}^t dt'\, G(t,t') C^{p-2}(t,t') \phi(t')
 + b(t) - z(t)\phi(t)
+ \eta(t)\,,
\intertext{where the effective reduced field is $b(t) = \beta J_0
M^{r-1}(t)$, the local response function is $G(t,t') = \delta \langle
\phi(t) \rangle / \delta b(t')$, the correlation function is $C(t,t')
= \langle \phi(t) \phi(t') \rangle$, the magnetization is $M(t) =
\langle \phi(t) \rangle$, and there is a renormalized Gaussian noise
$\eta(t)$ with zero mean and satisfying} \langle \eta(t) \eta(t')
\rangle &= 2 \delta(t-t') + \half p \beta^2 J^2 C^{p-1}(t,t')\:.
\end{align}
This equation cannot be solved exactly, but it is possible to obtain
self-consistent equations using the standard aging
assumption\cite{ck:agingprsg1}
\begin{gather}
\begin{split}
C(t,t') &= C_{\mathrm{st}}(t-t') + {\cal C}(t'/t)\,,
\\
G(t,t') &= G_{\mathrm{st}}(t-t') + \frac{1}{t} {\cal G}(t'/t)\,,
\end{split}
\end{gather}
taking the limit of long times and setting all the time derivatives to
zero. The calculation follows \cite{chs:dyn} and, as expected, gives
the equations (\ref{eqnM}), (\ref{eqn1q}), (\ref{eqn2q}), and
(\ref{MS}) obtained in the MS version of the replica analysis,
identifying
\begin{gather}
\begin{aligned}
{\cal C}(1) &= q_1\,, & {\cal C}(0) &= q_0\,,
\\
C_{\mathrm{st}}(0) &= 1-q_1\,, \qquad\qquad & C_{\mathrm{st}}(\infty) &= 0\,,
\end{aligned}
\end{gather}
and $M = M(\infty)$. The r\^ole played by $x$ is that of a factor in a
modified fluctuation-dissipation theorem (QFDT) in the non-ergodic phase:
\begin{gather}
\frac{1}{t}{\cal G}(t'/t) = - x\beta\, \Theta(t'/t) \frac{\partial
{\cal C}(t'/t)}
{\partial t'}.
\end{gather}
There is a direct correspondence between the breaking of replica symmetry
and the breaking of ergodicity.

\section{Results and interpretation}

\begin{figure}
\epsfig{file=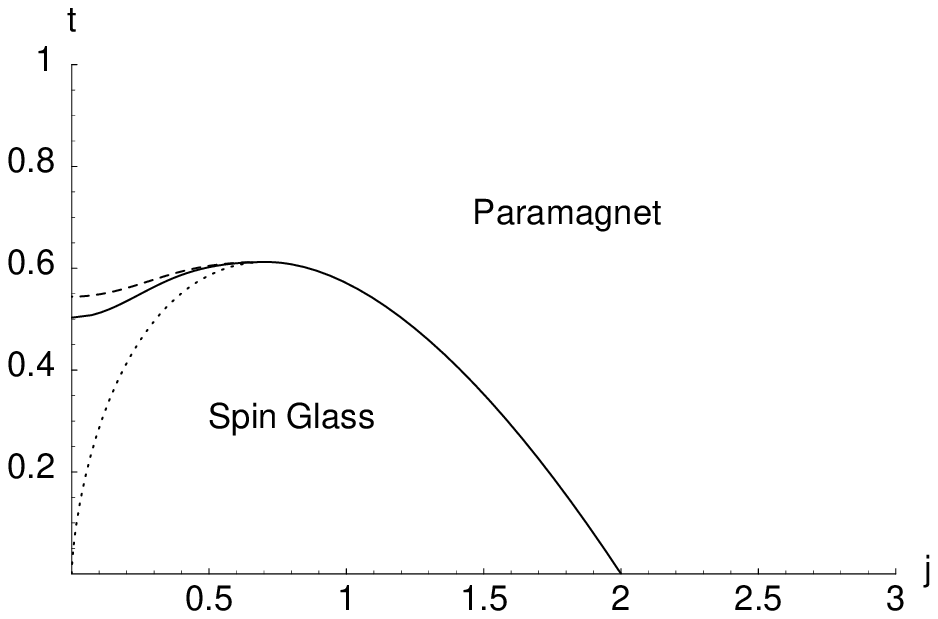}
\caption{The phase diagram the spin glass with $p=4$ in a magnetic
field $h$\cite{cs:stat,chs:dyn}. The axes are $j=h/J$ and $t=T/J$. The
static result is shown by the solid line; where the dynamic result
differs it is shown by the dashed line. The transitions are D1RSB to
the left of the maximum and C1RSB to the right. The dotted line is the
continuation of the Almeida--Thouless stability line where it does not
coincide with the C1RSB. For $p=4$, the static and dynamic transition
temperatures are $\Tsz \approx 0.503\, J$ and $T_\mathrm{d}^0 \approx
0.544\,J$ at $h = 0$, and both peak at $T = \Tm \approx 0.612\,J$, $h
= \Hm = J/\sqrt{2}$ and fall to zero at $h = \Hc = 2\,J$.}
\label{r1}
\end{figure}

\begin{figure}
\epsfig{file=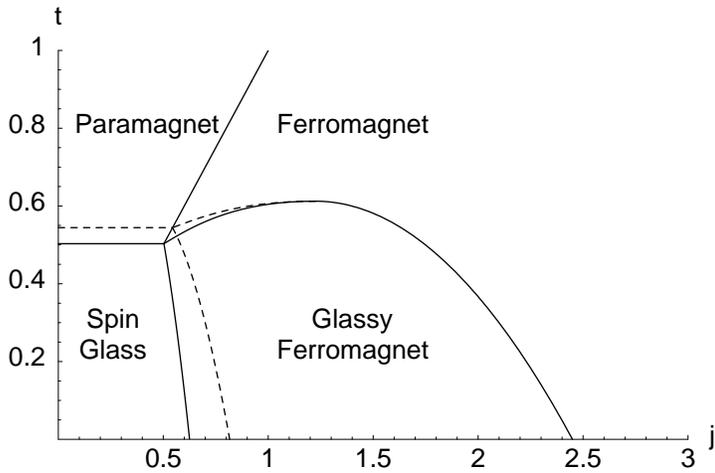}
\caption{The phase diagram for $p=4$ and $r=2$.\cite{hsn:r2} The axes are
$j=J_0/J$ and $t=T/J$. The static results are shown by the solid lines;
where the dynamic results differ they are shown by the dashed
lines. Transitions to glassy behaviour are D1RSB to the left of the
maximum and C1RSB to the right.}
\label{r2}
\end{figure}

\begin{figure}
\epsfig{file=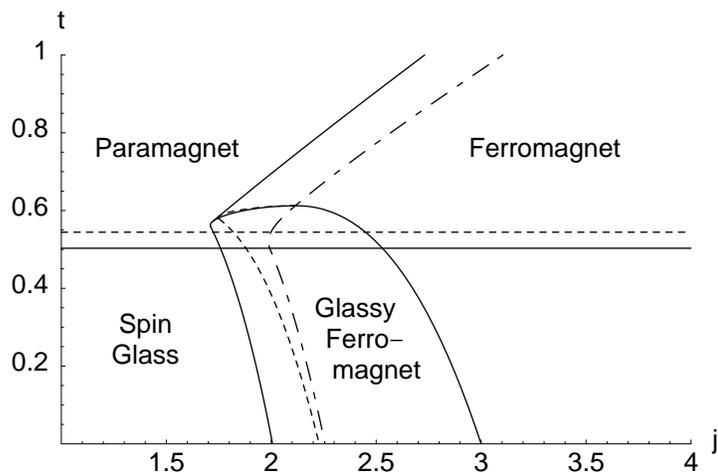}
\caption{The phase diagram for $p=4$ and $r=3$. The axes are $j=J_0/J$ and
$t=T/J$. The static spinodal results are shown by the solid lines; where
the dynamic results differ they are shown by the dashed line. The 
dot-dashed line shows the thermodynamic transitions. Transitions to
glassy behaviour are D1RSB to the left of the maximum and C1RSB to the
right.}
\label{r3}
\end{figure}

\begin{figure}
\epsfig{file=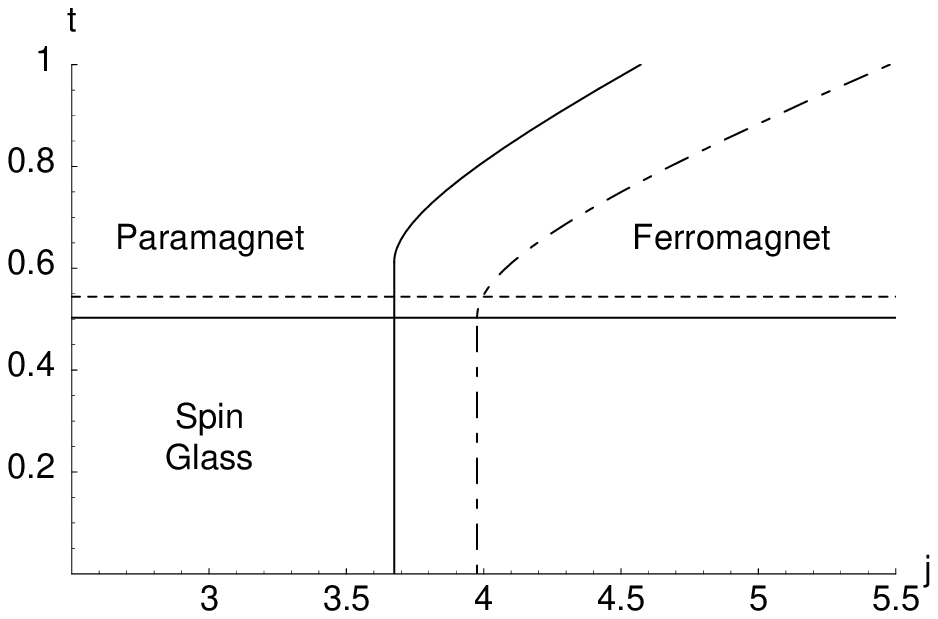}
\caption{The phase diagram for $p=r=4$. The axes are $j=J_0/J$ and
$t=T/J$. The static spinodal results are shown by the solid lines; where
the dynamic results differ they are shown by the dashed line. The 
dot-dashed line shows the thermodynamic transitions.}
\label{r4}
\end{figure}

\begin{figure}
\epsfig{file=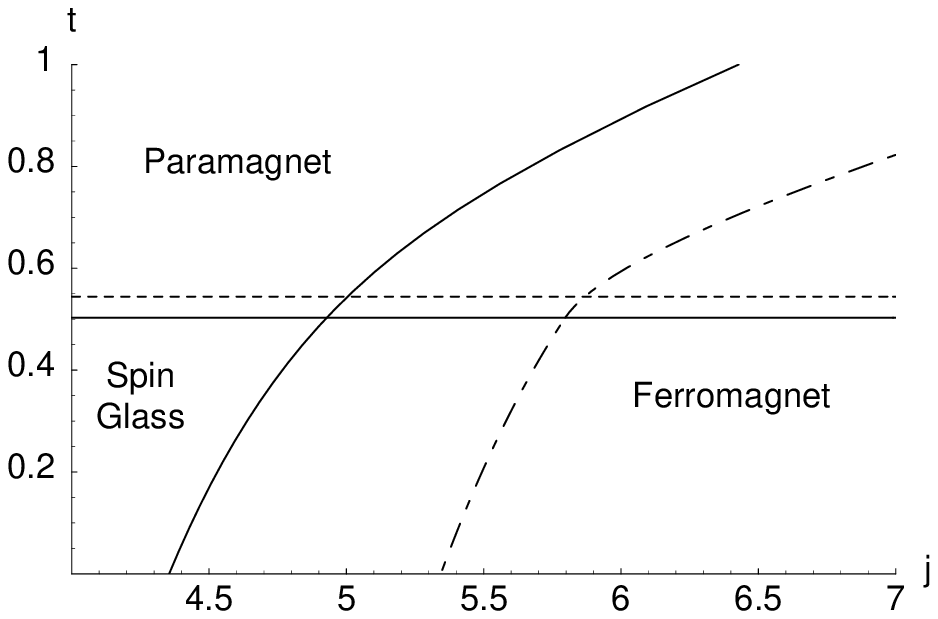}
\caption{The phase diagram for $p=4$ and $r=5$. The axes are 
$j=J_0/J$ and $t=T/J$. The static spinodal results are shown by
the solid lines; where the dynamic results differ they are shown
by the dashed line. The dot-dashed line shows the thermodynamic
transitions.}
\label{r5}
\end{figure}

Phase diagrams for a characteristic set of situations are exhibited in
Figures~\ref{r1}--\ref{r5}; the new results are in
Figures~\ref{r3}--\ref{r5} but Figures~\ref{r1} and \ref{r2} are
included for orientation (as well as completeness). In each case $p=4$
is chosen, but similar results apply for other $p \geqslant 3$. The
remaining figures show the free energies of systems with constrained
magnetizations: these assist in the interpretation of the phase
diagrams which result when the constraint is removed and the free
energy minimized with respect to $M$.

Figure~\ref{r1} is for $r=1$ \cite{cs:stat,chs:dyn}, in which case the
second term of (\ref{ham}) corresponds to an applied field $h=J_0$. It
is helpful to recall its features. For $h>\Hc$ the only stable state
is paramagnetic. For $h$ between $\Hc$ and $\Hm$ there is a
continuous one-step replica symmetry breaking (C1RSB) transition from
paramagnet to spin glass, with $(q_1-q_0) \to 0 $ at the transition;
the transition temperature is the same statically and dynamically, and
the transition coincides with the onset of Almeida--Thouless
instability. At $h=\Hm$, this temperature reaches a maximum of
$\Tm$. For $h<\Hm$ there is a discontinuous one-step replica
symmetry breaking (D1RSB) transition from paramagnet to spin glass,
with $x \to 1$ at the transition. The transition temperature is higher
for dynamics (or equivalently marginal stability) than for statics;
both transition temperatures are higher than that at which small
fluctuation Almeida--Thouless instability would onset were it not
preempted by the discontinuous transition. All three temperatures,
which we shall label $\Td$, $\Ts$, and $\Tat$ respectively, fall as
$h$ falls; for future use we define $\Ts=\Tsz$ at $h=0$. In
considering the various systems with ferromagnetic interactions
discussed below it will be helpful to make reference to this case.

\begin{figure}
\subfigure[$T/J=0.7$, $J_0/J=0$]
{\epsfig{file=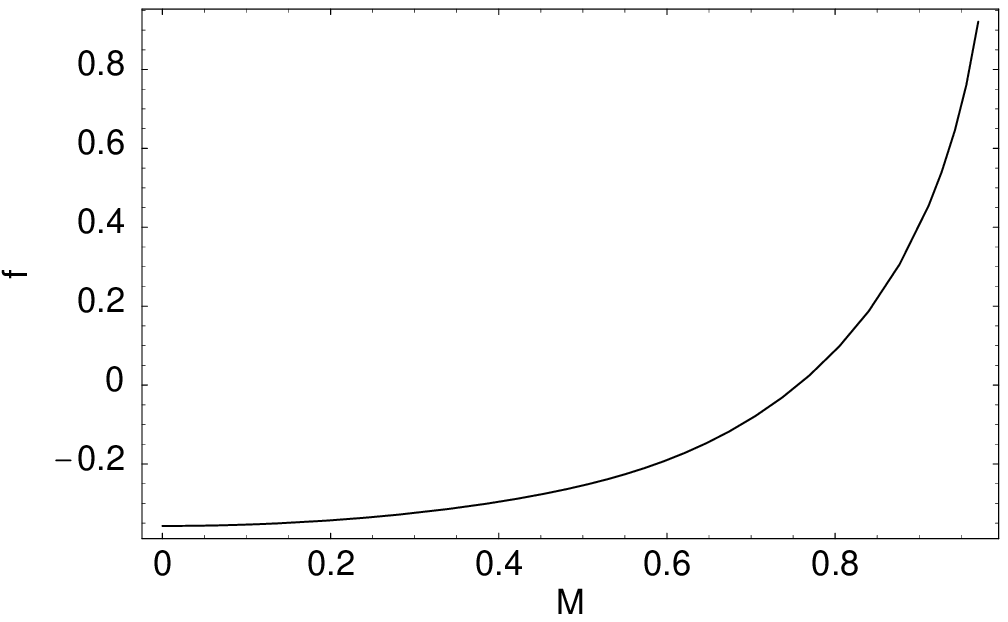,width=2.4in}}
\qquad
\subfigure[$T/J=0.505$, $J_0=0$]
{\epsfig{file=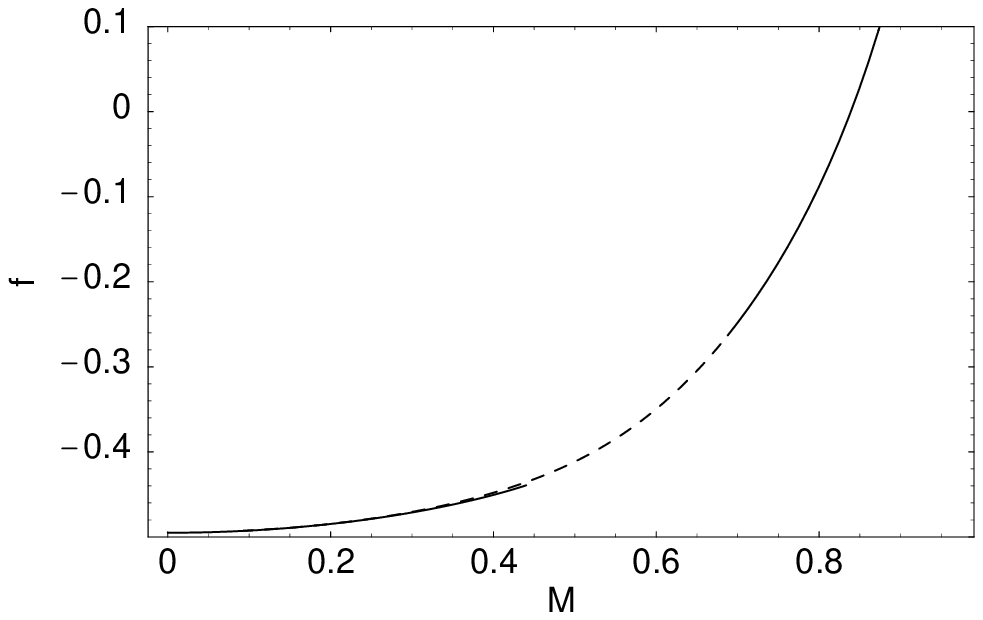,width=2.4in}}
\\
\subfigure[$T/J=0.505$, $J_0=0$ (close-up)]
{\epsfig{file=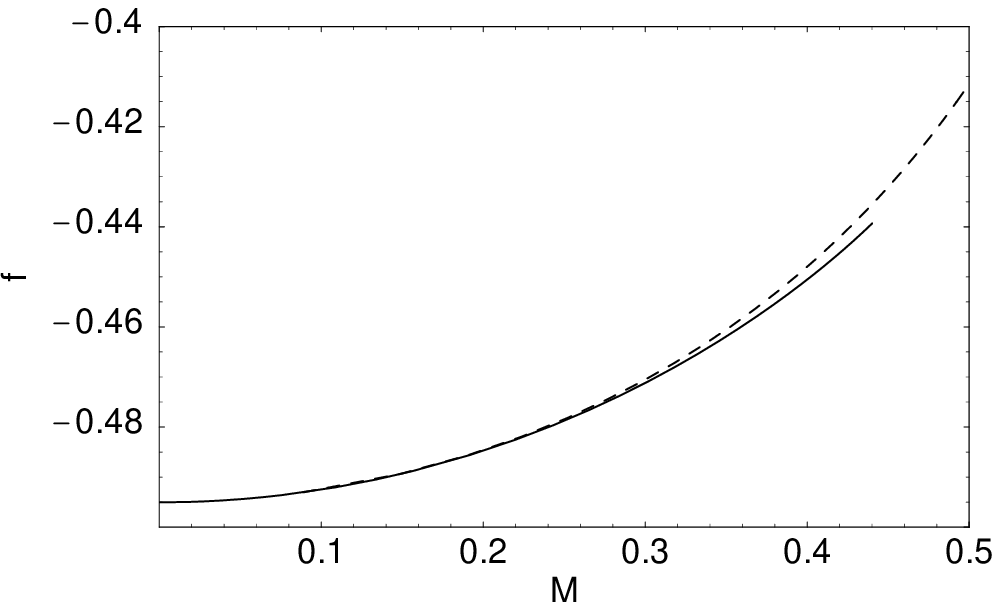,width=2.4in}}
\qquad
\subfigure[$T/J=0.45$, $J_0=0$]
{\epsfig{file=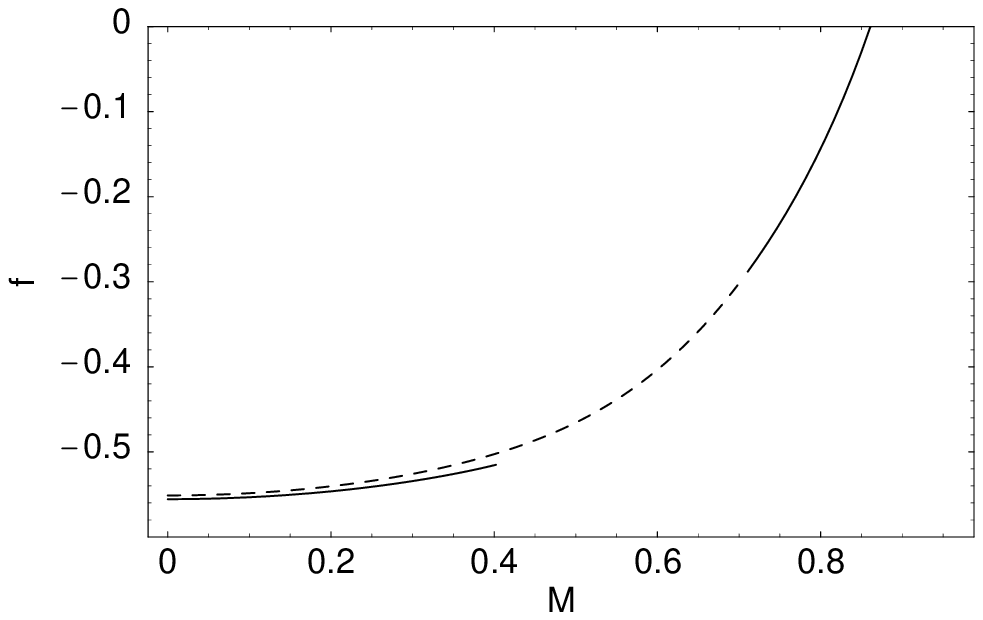,width=2.4in}}
\caption{Plots of the free energy per site $f$ against the constrained
magnetization $M$ at various temperatures when $J_0=0$. (In this case
there is no dependence on $r$.) The solid lines give the RS solutions,
the dashed lines the RSB.}
\label{Mj0}
\end{figure}

Before passing to the generalized models it is also useful to consider
the system with $J_0=0$ (for which case the value of $r$ is
irrelevant) but with constrained magnetization. For $T>\Tm$ the free
energy as a function of $M$ has the form shown in Figure~\ref{Mj0}(a):
the stable state is replica symmetric for all $M$. As $T$ is reduced
below $\Tm$ the character of $f(M)$ changes as shown in
Figure~\ref{Mj0}(b), or in more detail in Figure~\ref{Mj0}(c): a gap
opens up in which there is no longer an RS solution stable against
Almeida--Thouless fluctuations, and a region in which there is a new
RSB solution appears. The RSB solution spans the gap in the RS
solution, with an overlap at its lower end. That is, the upper
end-point of the RSB curve coincides with the lower end-point of the
high-$M$ section of the RS curve, but the lower end-point of the RSB
curve lies on the low-$M$ section of the RS curve below its
end-point. The RSB solution has monotonically increasing $(q_1-q_0)$
as $M$ is lowered below the upper connection point of RS and RSB. The
coincidence at the upper end of the gap is related to the possibility
of a continuous RSB, while the overlap at the lower end is related to
a discontinuous RSB. As the temperature is lowered further the gap in
the RS curve and range of RSB both grow, with the latter extending to
$M=0$ at $T \leqslant \Tsz$, as shown in Figure~\ref{Mj0}(d). For $J_0=0$
the minimum of $f(M)$ is always at $M=0$ which is therefore the
unconstrained magnetization. For $T<\Tsz$, where both RS and RSB
solutions exist at $M=0$, the latter is favoured.

Increasing $h=J_0$ for $r=1$ modifies the curves $f(M)$ and moves the
minimum to a finite magnetization $M=\Mm$. For $T>\Tm$, $f(M)$ remains
only RS and the unconstrained state remains paramagnetic. For
$\Tsz<T<\Tm$, where there is an RSB curve which does not extend to
$M=0$, the sequence of events on increasing $h$ is as follows:
(i)~the minimum moves out along the lower section of the RS curve,
(ii)~it crosses into the region where the RS and RSB curves overlap
and both have minima, the RSB being favoured so that a D1RSB
transition to a spin glass takes place; (iii)~the RSB minimum
continues to move out, while the RS minimum reaches the end of the
lower RS branch and disappears, corresponding to crossing the
Almeida--Thouless line, but the RS minimum is irrelevant and no phase
change occurs; (iv)~the RSB minimum moves up the RSB curve until this
gives way smoothly to the upper section of the RS curve and a C1RSB
transition back to a paramagnet takes place. For $T<\Tm$ again the RSB
curve does extend to $M=0$ and the system already favours the spin
glass solution at $h=0$, so only (iii) and (iv) occur.

For $r \geqslant 2$ ferromagnetism becomes possible with effective field
\begin{gather}
\he = J_0 M^{r-1}
\end{gather}
determined self-consistently. Figure~\ref{r2} shows the phase diagram
obtained recently \cite{hsn:r2} for $r=2$. At low $J_0$ the
frustration due to the disorder in $J$ continues to prevent
ferromagnetism, as does entropy as the temperature is raised, leading
to behaviour similar to that for $J_0=0$. At a temperature-dependent
$J_0$ a continuous ferromagnetic transition takes place and the system
goes over to a solution whose magnetization rises continuously with
$J_0$. The ferromagnetic region is split into two parts, non-glassy RS
(at higher $J_0$, $T$) and glassy 1RSB (at lower $J_0$, $T$). The
transition between them directly mirrors the behaviour of
Figure~\ref{r1} with $h$ replaced by the self-consistently determined
$\he$ and is D1RSB (C1RSB) for $J_0$ less (greater) than the value
$\Jm$ for which $\he=\Hm$. (The transition temperature reaches the
maximum of $\Tm$ at $J_0=\Jm$.) Again there is an Almeida--Thouless
curve which lies beneath the D1RSB transition line for $J_0<\Jm$ but
is coincident with the C1RSB transition line for $\Jm<J_0<\Jc$ where
$\Jc$ is the value of $J_0$ at which $\he=\Hc$.

\begin{figure}
\subfigure[$r=2$, $T/J=0.7$, $J_0/J=1$]
{\epsfig{file=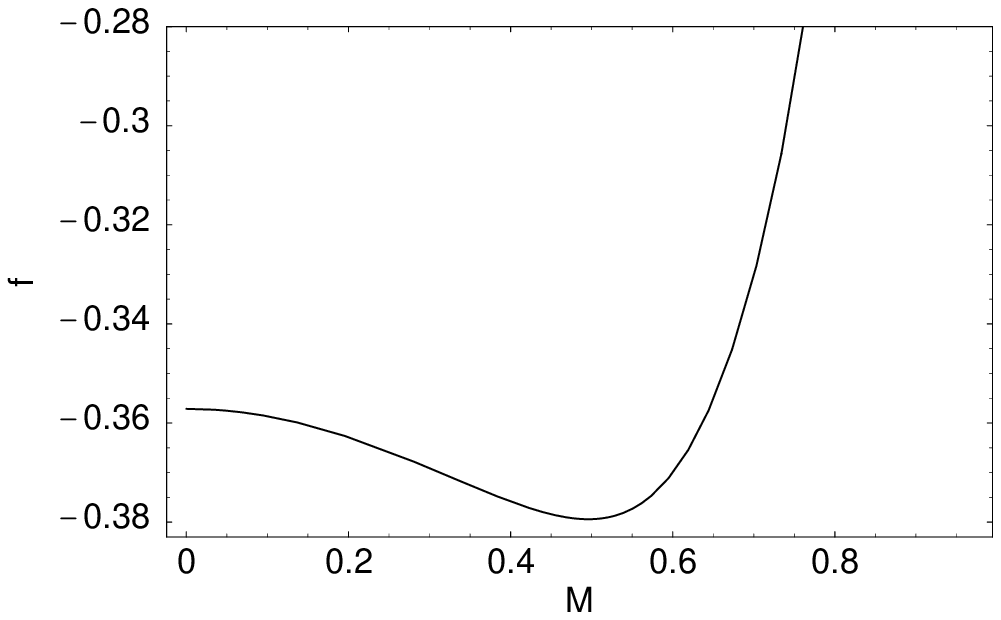,width=2.4in}}
\qquad
\subfigure[$r=2$, $T/J=0.55$, $J_0/J=0.6$]
{\epsfig{file=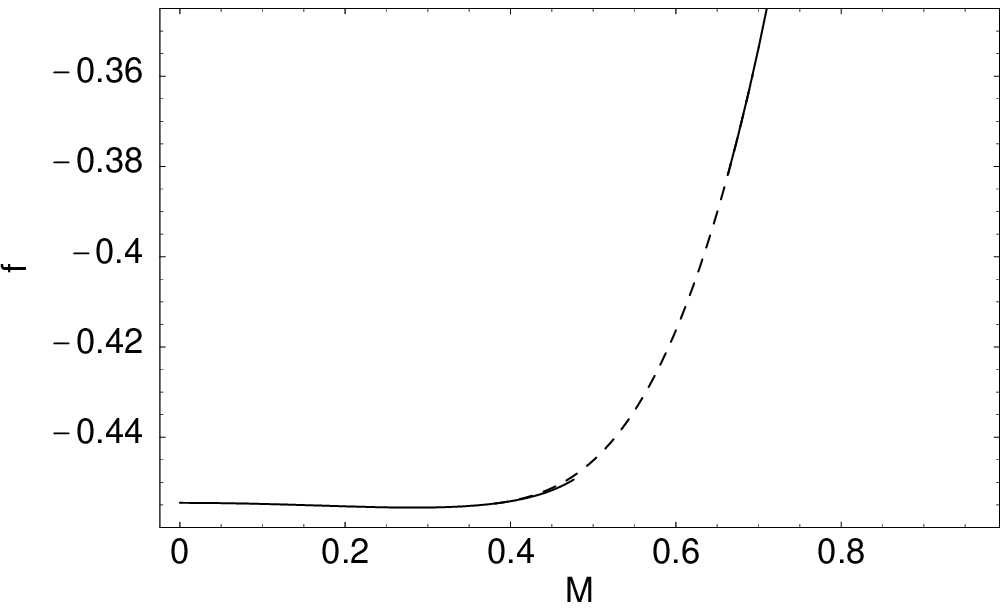,width=2.4in}}
\\
\subfigure[$r=2$, $T/J=0.55$, $J_0/J=1.3$]
{\epsfig{file=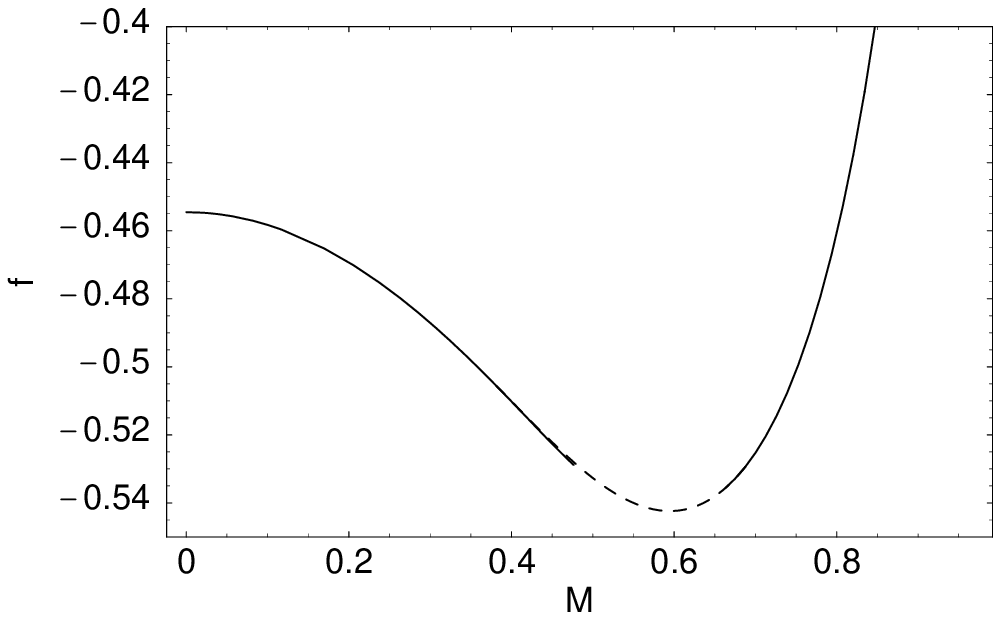,width=2.4in}}
\qquad
\subfigure[$r=2$, $T/J=0.55$, $J_0/J=2.4$]
{\epsfig{file=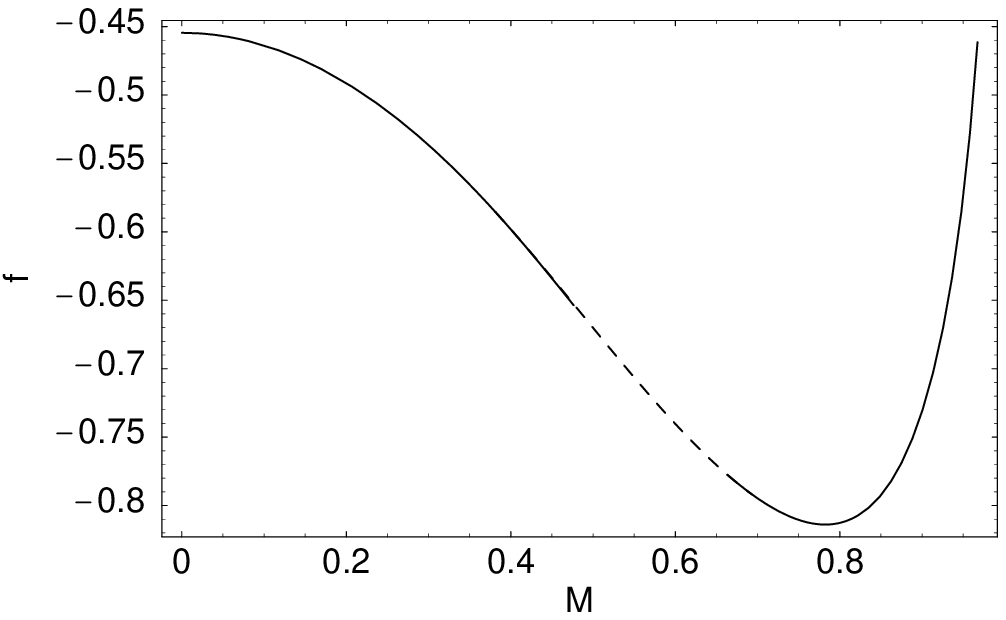,width=2.4in}}
\caption{Plots of the free energy per site $f$ against the constrained
magnetization $M$ at various points in the phase diagram when $r=2$. The
solid lines give the RS solutions, the dashed lines the RSB.}
\label{Mr2}
\end{figure}

The transitions are apparent in $f(M)$ through behaviour directly
analogous to that discussed above for $r=1$. In the case of $r=2$ the
ferromagnetic onsets are second order, with the minimum in $f(M)$
moving away from $M=0$ continuously as $J_0$ is increased across the
transition lines. Figure~\ref{Mr2} illustrates several aspects of the
phase diagram: \ref{Mr2}(a) shows a region of non-glassy
ferromagnetism above $\Tm$, where $f(M)$ is RS throughout its range;
\ref{Mr2}(b) shows a non-glassy ferromagnet between $\Tm$ and $\Tsz$
for $J_0<\Jm$, with a gap in the RS curve and an RSB section but with
the minimum in the lower RS region; \ref{Mr2}(c) shows a glassy
ferromagnet at the same temperature, where the minimum is now in the
RSB region, the system having undergone a D1RSB transition;
\ref{Mr2}(d) shows a non-glassy ferromagnet at the same temperature
for $J_0>\Jm$, with the minimum now in the upper RS region, the system
having undergone a C1RSB transition.

\begin{figure}
\subfigure[$r=3$, $T/J=0.7$, $J_0/J=2.1$]
{\epsfig{file=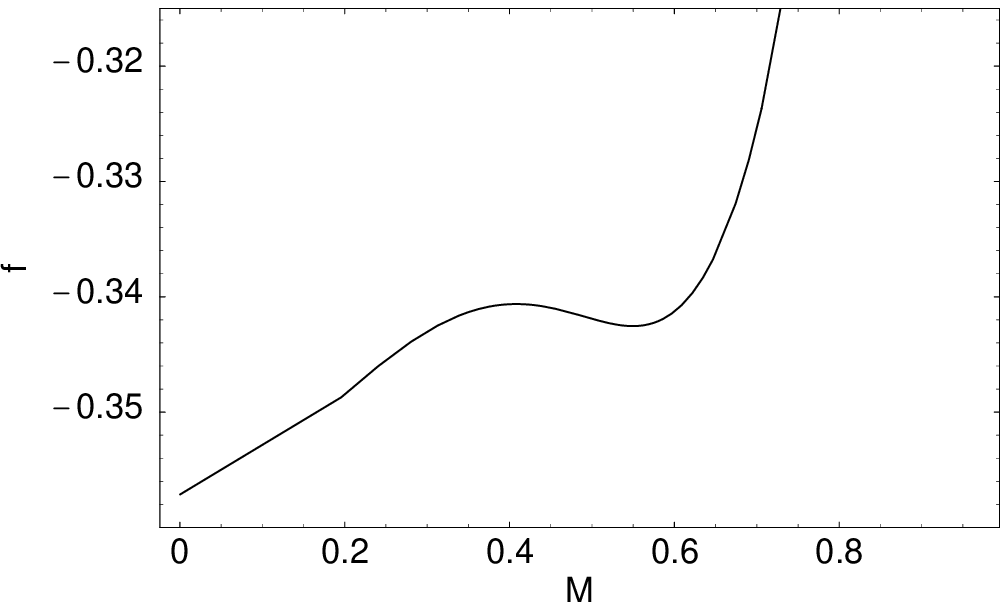,width=2.4in}}
\qquad
\subfigure[$r=3$, $T/J=0.7$, $J_0/J=2.5$]
{\epsfig{file=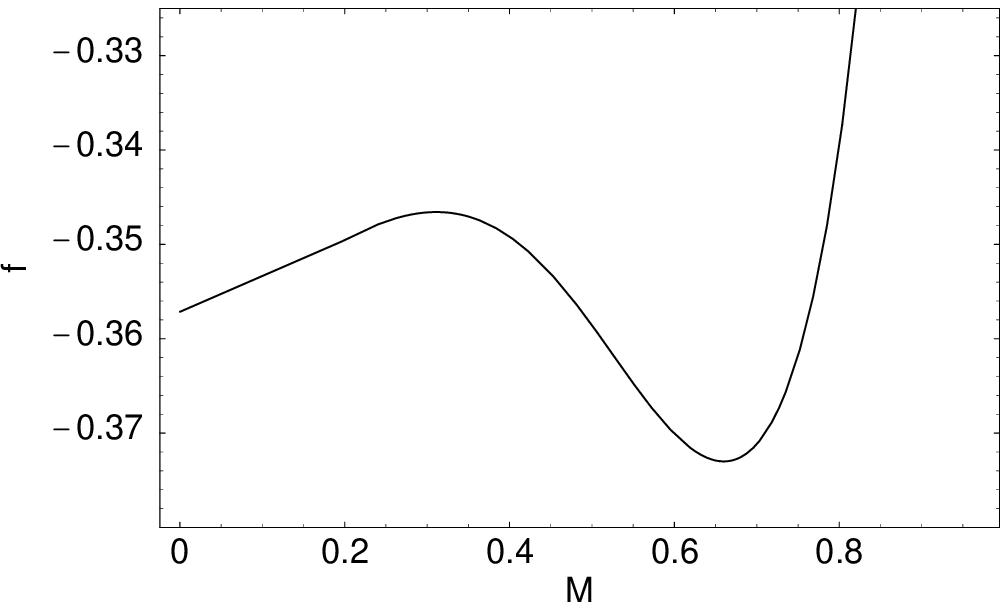,width=2.4in}}
\\
\subfigure[$r=3$, $T/J=0.3$, $J_0/J=1.8$]
{\epsfig{file=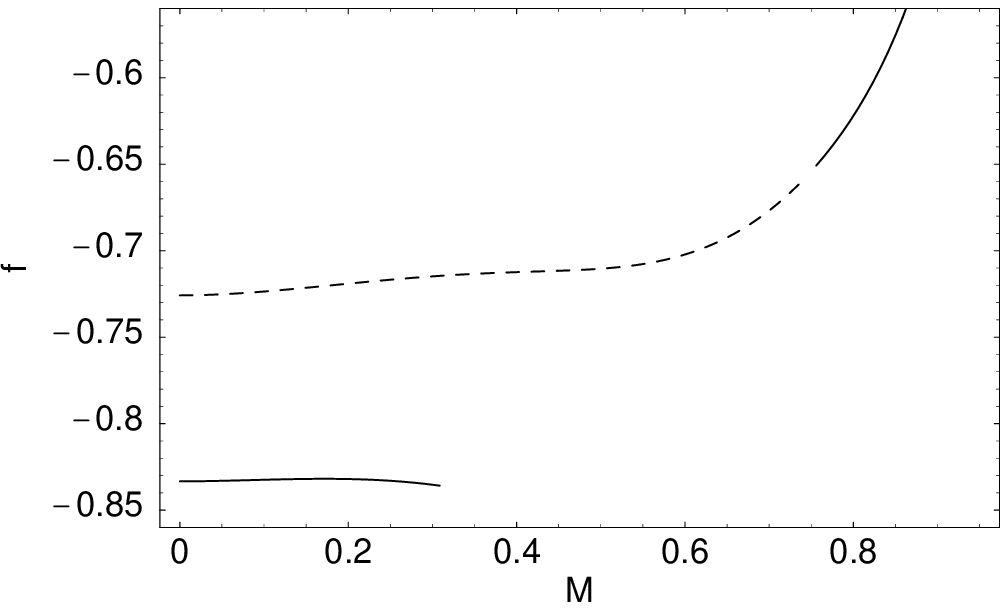,width=2.4in}}
\qquad
\subfigure[$r=3$, $T/J=0.3$, $J_0/J=2$]
{\epsfig{file=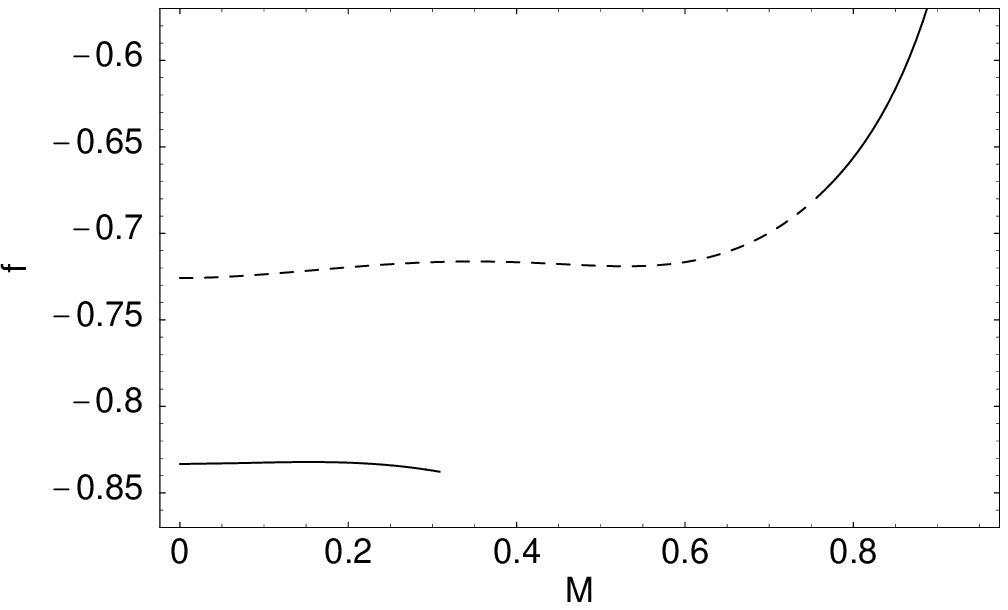,width=2.4in}}
\\
\subfigure[$r=3$, $T/J=0.3$, $J_0/J=2.5$]
{\epsfig{file=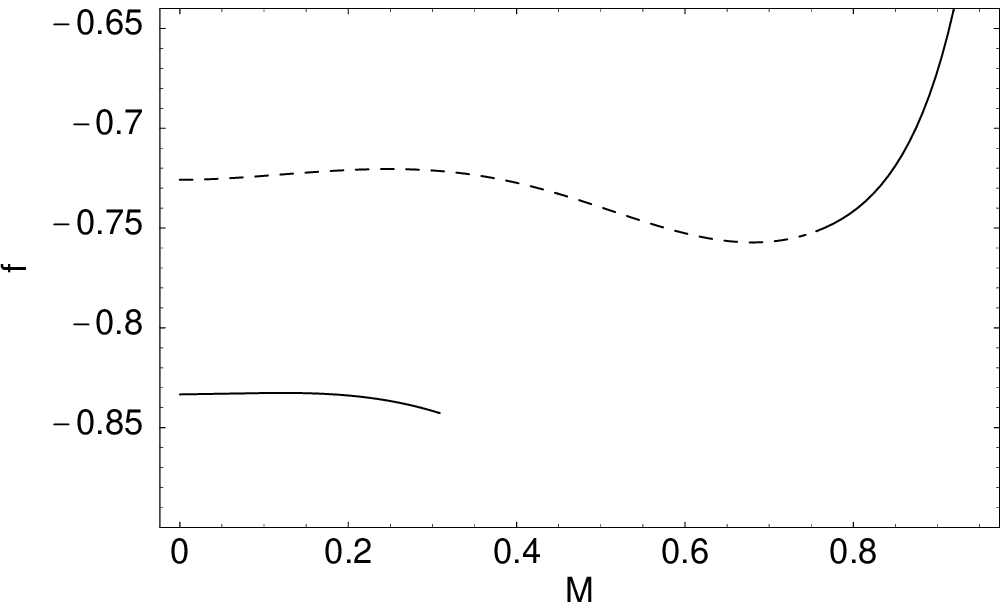,width=2.4in}}
\qquad
\subfigure[$r=3$, $T/J=0.3$, $J_0/J=3$]
{\epsfig{file=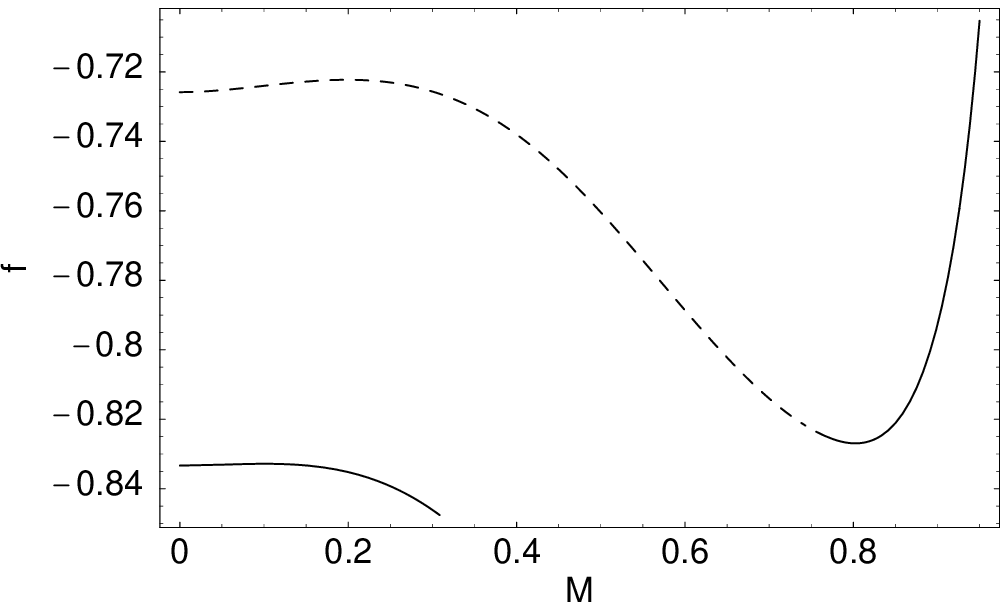,width=2.4in}}
\caption{Plots of the free energy per site $f$ against the constrained
magnetization $M$ at various points in the phase diagram when $r=3$. The
solid lines give the RS solutions, the dashed lines the RSB.}
\label{Mr3}
\end{figure}

\begin{figure}
\subfigure[$r=4$, $T/J=0.3$, $J_0/J=3.7$]
{\epsfig{file=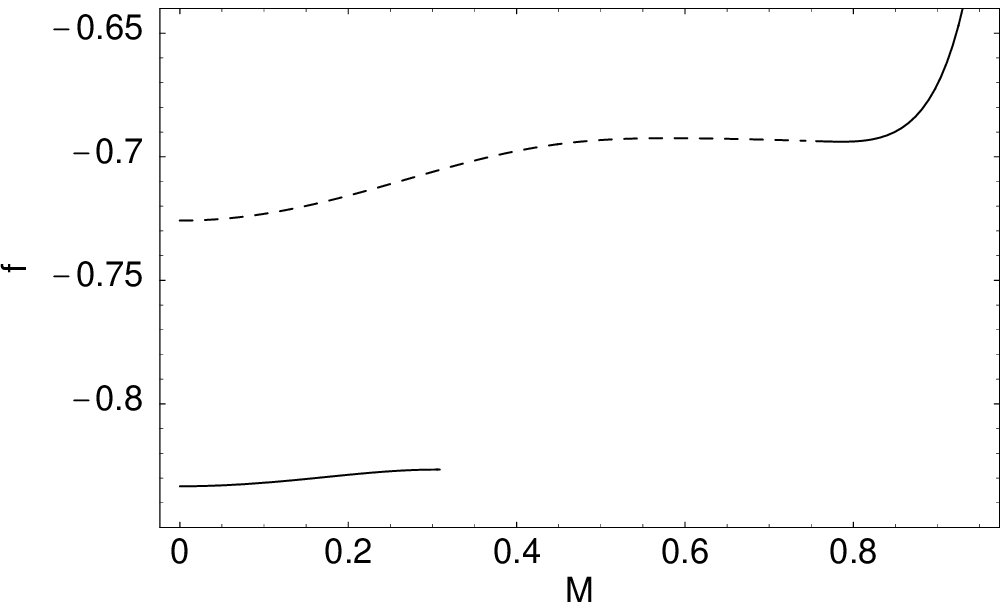,width=2.4in}}
\qquad
\subfigure[$r=5$, $T/J=0.3$, $J_0/J=4.7$]
{\epsfig{file=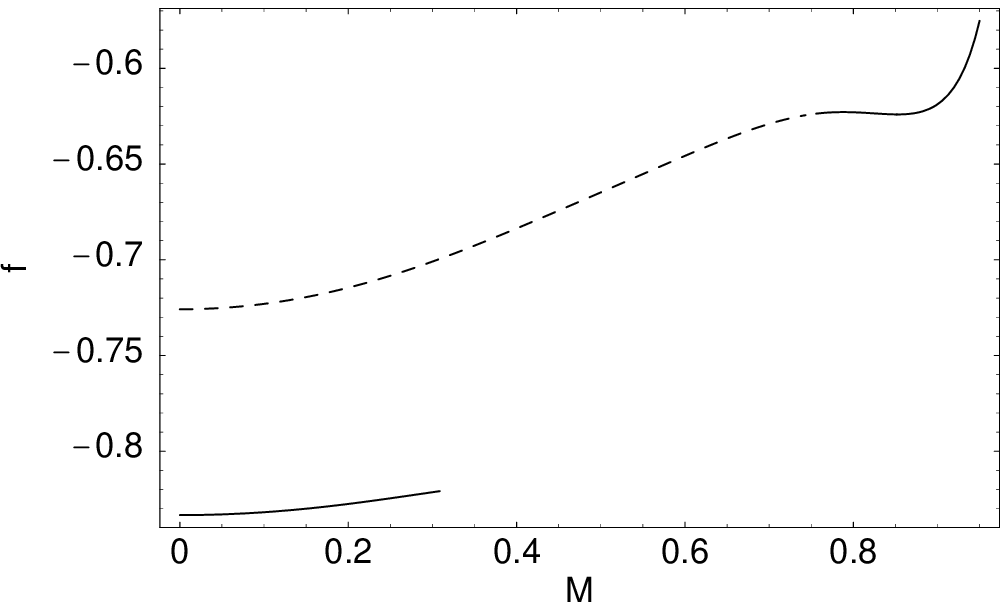,width=2.4in}}
\caption{Plots of the free energy per site $f$ against the constrained
magnetization $M$ at a point near the spinodal transition between spin
glass and ferromagnet for $r=4$ and $r=5$. The solid lines give the RS
solutions, the dashed lines the RSB.}
\label{Mr45}
\end{figure}

For $r>2$ the ferromagnetic transitions are first order, with the
$M=0$ solution always locally stable. As noted before, there are two
kinds of transition as $J_0$ is increased or $T$ is decreased: a
spinodal transition at which a secondary minimum appears in $f(M)$ at
a finite $M$ whilst the lowest minimum is at $M=0$, and a
thermodynamic transition at which the finite $M$ minimum becomes lower
than that at $M=0$. Figures~\ref{r3}--\ref{r5} show the full phase
diagrams for $p=4$ and $r=3,4,5$ respectively, as characteristic
illustrations of systems with first order transitions for which $r$ is
less than, equal to, and greater than $p$. Only for the case $r<p$ is
a glassy ferromagnet found with glassy/non-glassy transitions; this
transition is analogous to that for the case $r=2$, with D1RSB for
$J_0$ less than $\Jm$ and C1RSB for $J_0>\Jm$. Figure~\ref{Mr3}
illustrates the underlying character of $f(M)$, which drives the
static transitions. Figures~\ref{Mr3}(a,b) show the situation for
$T>\Tm$, (a) in a region of spinodal ferromagnetism, (b) of
thermodynamic ferromagnetism. Figures~\ref{Mr3}(c--f) are for
$T<\Tsz$, showing the effect of increasing $J_0$: (c) is in the spin
glass phase, passing to an RSB ferromagnet (d) via a spinodal
transition, with the RSB ferromagnet thermodynamically preferable to
the spin glass in (e), and eventually crossing via a continuous
transition into an RS ferromagnet in (f). The smallest self-consistent
value of $\he$ (i.e. that at the spinodal transition) increases with
$r$, and in fact for $r=p$ at any $T<\Tm$ it is exactly that
corresponding to the C1RSB line in Figure~\ref{r1} so that the
transition is directly into an RS ferromagnet. This is shown in
Figure~\ref{Mr45}(a). (Note that since the C1RSB line coincides with
the Almeida--Thouless line, the RS ferromagnet does become unstable
against RSB fluctuations at the transition.) For $r>p$ the smallest
$\he$ is beyond this critical value and carries the system well into
the RS region, so there is no trace of an RSB ferromagnet. This is
shown in Figure~\ref{Mr45}(b). Hence for all $r \geqslant p$ glassy
ferromagnetism is suppressed. In the case of $r=p$ the minimum $\he$
corresponds to the same value of $J_0$ for any $T<\Tm$, yielding a
vertical transition line between spin-glass and RS ferromagnet.

\begin{figure}
\epsfig{file=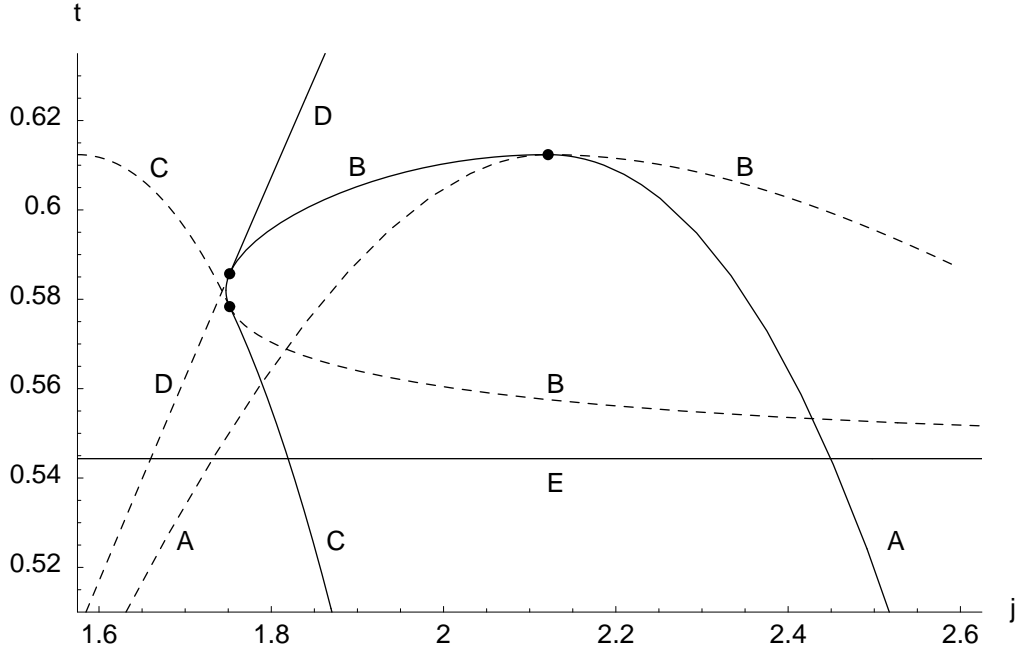}
\caption{A detail of the dynamic spinodal phase diagram for $p=4$ and
$r=3$ shown in Figure~\ref{r3}. The axes are $j=J_0/J$ and
$t=T/J$. The solid lines are the actual phase transition lines, the
dashed lines are their continuations into regions where they are
superseded by earlier transitions. For clarity, the points where the
various curves meet are marked with dots. A qualitatively similar
figure applies for the static spinodal lines. The onset of the glassy
ferromagnet is given by 1RSB solutions to (\ref{eqnM}), (\ref{eqn1q}),
(\ref{eqn2q}), and (\ref{eqnx}), with $M \neq 0$ and the additional
constraints $q_1=q_0$ on (A), $x=1$ on (B), and $\dFRSB=0$ on (C). The
onset of the non-glassy ferromagnet is given by RS solutions to the
equations with $M \neq 0$ and $\dFRS=0$, and is shown by (D). The
onset of the spin glass is given by 1RSB solutions to the equations
with $M=0$ and $x=1$, and is shown by (E).}
\label{r3spin}
\end{figure}

For $r>p$ (as in Figure~\ref{r5}) the ferromagnet becomes marginally
stable against small fluctuations in $M$ along its spinodal transition
line $\dFRS=0$. The corresponding thermodynamic transition line is in
two segments: for $T>\Tsz$ it is given by $\FFM=\FPM$; whilst for
$T<\Tsz$ it is where $\FFM=\FSG$. For $r=p$ (as in Figure~\ref{r4})
the spinodal transition is also in two segments: for both segments it
has $\dFRS=0$ as before, but in the lower section ($T<\Tm$) it
simultaneously becomes unstable against Almeida--Thouless
fluctuations. The thermodynamic transition is as for $r>p$. For $r<p$
(as in Figure~\ref{r3}) the spinodal transition line to glassy
ferromagnetism is in three segments (see Figure~\ref{r3spin}): (A) the
C1RSB line where $q_1=q_0$ and $x \neq 1$; (B) the D1RSB line where
$x=1$ and $q_1 \neq q_0$; (C) the line $\dFRSB=0$, where it is
marginally stable against small fluctuations in $M$ again. Non-glassy
ferromagnetism onsets at $\dFRS=0$, as shown by curve (D) which
terminates where it intersects (B). The static and dynamic results are
qualitatively the same, although the lines (B) and (C) are slightly
displaced. The thermodynamic transition lines are as the spinodal
lines for the continuous transitions, but for the first order
ferromagnetic transitions are moved to higher $J_0$. This curve is in
three segments: (C) has moved to the line $\FSG=\FGFM$; the lower
section of (B) to $\FPM=\FGFM$; (D) has moved to $\FPM=\FFM$.

\section{Conclusions}
In this paper we have solved the infinite range spherical $p$-spin
glass with an additional $r$-spin ferromagnetic interaction, finding
the phase diagrams both in statics and dynamics for $p>2$ and general
$r$. By further examination of the free energy with constrained
magnetization we have clarified the origin of the different phases,
both for the previously studied models with $r=1,2$ and for $r>2$. We
have related the behaviour of systems with $r \geqslant 2$ to those with
$r=1$. In all cases, in the replica method, the first step of replica
symmetry breaking is sufficient. As previously noted \cite{hsn:r2},
for $r=2$ there are thermodynamically continuous transitions to
ferromagnetism with two types of ferromagnetic region, non-glassy and
glassy (Figure~\ref{r2}). For $r>2$ the ferromagnet transitions are
first order. For $r<p$ there remains a region of parameter space where
the system favours a glassy ferromagnet (Figure~\ref{r3}), whereas for
$r \geqslant p$ glassy ferromagnetism is suppressed and the ferromagnetic
region consists of a single non-glassy phase
(Figures~\ref{r4},~\ref{r5}).

\section*{Acknowledgments}
The authors would like to thank Andrea Cavagna, Irene Giardina,
Juan-Pedro Garrahan and John Hertz for helpful discussions. We would
also like to thank EPRSC (UK) for financial support, PG for research
studentship 97304251 and DS for research grant GR/M04426.

\newpage

\end{document}